\documentclass{acm_proc_article-sp}
\usepackage{hyperref}
\hypersetup{
    bookmarks=false,         
    unicode=false,          
    pdftoolbar=true,        
    pdfmenubar=true,        
    pdffitwindow=false,     
    pdfstartview={FitH},    
    pdftitle={My title},    
    pdfauthor={Author},     
    pdfsubject={Subject},   
    pdfcreator={Creator},   
    pdfproducer={Producer}, 
    pdfkeywords={keyword1} {key2} {key3}, 
    pdfnewwindow=true,      
    colorlinks=true,       
    linkcolor=black,          
    citecolor=black,        
    filecolor=black,      
    urlcolor=black           
}

\usepackage{makeidx}
\usepackage{graphicx}
\usepackage{amssymb}
\usepackage{amsmath}
\usepackage{amscd}
\usepackage{tikz}
\usetikzlibrary{matrix,arrows}
\usepackage{multicol}

\newcommand{\comment}[1]{}
\definecolor{Orange}{rgb}{1,0.5,0}

\newcommand{\commentout}[1]{}

\newcommand{\R}{\mathbb{R}}                    

\newcommand{\args}[1]{\mathop{\left( #1 \right)}}

\newcommand{\cbrace}[1]{\mathop{\left\{ #1 \right\}}}

\newcommand{\argsS}[2]{\mathop{\left( #1 \right)#2}}

\renewcommand{\S}[1]{{\mathcal{#1}}}           
\def\vec#1{\mathchoice{\mbox{\boldmath$\displaystyle#1$}}
{\mbox{\boldmath$\textstyle#1$}}
{\mbox{\boldmath$\scriptstyle#1$}}
{\mbox{\boldmath$\scriptscriptstyle#1$}}}

{%
\end{array}\right.}

\newcounter{algorithm_counter}
{
\\[-1.5ex]
\rule{\textwidth}{\arrayrulewidth}
\end{footnotesize}
\end{minipage}
\setlength{\parindent}{\parindent}
}

{
\\[-1.5ex]
\rule{\textwidth}{\arrayrulewidth}
\end{footnotesize}
\end{minipage}
\setlength{\parindent}{\parindent}
}

\begin{document}

\title{The Gamification Design Problem}


%
%
%
%
%

\numberofauthors{2} 
%
\author{
%
%
\alignauthor
Michael Meder\\
      \affaddr{Technische Universit\"at Berlin}\\
      \email{meder@dai-lab.de}
\alignauthor
Brijnesh-Johannes Jain\\
      \affaddr{Technische Universit\"at Berlin}\\
      \email{jain@dai-lab.de}
}


\maketitle

\begin{abstract}
Under the assumptions that (i) gamification consists of various types of users that experience game design elements differently; and (ii)  gamification is deployed in order to achieve some goal in the broadest sense, we pose the gamification problem as that of assigning each user a game design element that maximizes their expected contribution in order to achieve that goal. We show that this problem reduces to a statistical learning problem and suggest matrix factorization as one solution when user interaction data is given. The hypothesis is that predictive models as intelligent tools for supporting users in decision-making may also have potential to support the design process in gamification.
\end{abstract}




\section{Introduction}
\label{sec:intro}
Since 2010 the application of gamification \cite{Deterding2011d, Huotari2012a} is a trending topic for marketing and business oriented services but also receives more and more recognition by the scientific community \cite{Hamari2014}. Gamification has been applied in various environments and for different purposes such as enterprise workplaces, education, pervasive health care, e-commerce, human resource management and many more (e.g. \cite{Anderson2013,Cheng2011a,Nikkila2011}). Although these studies indicate that gamification can lead to increased user activity, a detailed analysis of users' personal perception of gamification principles has barely been studied. 
Existing gamification definitions pursuing the increase of user experience \cite{Deterding2011d} and overall value \cite{Huotari2012a} indicate that the application of gamification is goal oriented. Therefore, we look at gamification as the necessity to \textbf{maximize an overall goal}.
\\
We are all individuals and are driven by different input factors such as our personality, as well as social or cultural differences \cite{Hamari2013a,Khaled2011,Yang2011,Yee2006,Yee2012}. For example in an enterprise scenario, it is of uttermost importance to measure challenges and risks that occur due to these differences. On the one hand, we expect gamification to increase user participation within an enterprise. On the other hand, the visibility of user interaction (or lack thereof), e.g., the position of the employee on a leaderboard can increase the stress level of employees or even cause fear that their activities on a gamified system will be used as an indicator of their engagement with the company. Various negative effects of gamification are thinkable and already observed \cite{Hamari2014, Mosca2012}. For a successful gamification several factors needs to be considered, what makes the implementation difficult and expensive.
Therefore, our extended look at gamification is the necessity to \textbf{maximize an overall goal} with \textbf{respect to the individuality of users}.
\\
For designing games, one method to approach this individuality is to regard well known player typologies \cite{Hamari2014a}, which grouping similar players, and design an implementation of gamification addressing all existing player types. 
The most common technique to find out the user types is the use of questionnaires and interviews. However, this approach is associated with high efforts. Given the bias effect caused by questionnaires, we arguing it is hard to conclude on users' actual behavior in a gamified environment \cite{Meder2013}. Recent studies also indicate that the effect of game design elements can change over time, which can end up with lower effects on long-term \cite{Hamari2013, Farzan2008, Farzan2008a}. In a worst case a positive effect might only be caused by the novelty effect.
\\
In this paper we propose a new approach to solve the gamification design problem. We suggest matrix factorization to create a generic model based on user interaction data as a suitable methodology which could help for the selection of most fitting game design elements.

\section{Gamification Design}
\label{sec:gamedesign}
When designing an application of gamification one will unavoidable be faced by the question which game design elements to choose for the implementation. In this section we give, based on existing definitions, a short introduction of how we understand gamification and what we argue as a solution to find a helpful answer of the above question.

\subsection{Gamification Definitions}
\label{sec:gamedesigndef}
The term gamification was coined in blog posts by Bret Terril\footnote{http://www.bretterrill.com/2008/06/my-coverage-of-lobby-of-social-gaming.html} and James Currier\footnote{http://blog.oogalabs.com/2008/11/05/gamification-game-mechanics-is-the-new-marketing/} in 2008. In 2010, the term was adopted by both industry and academia. In 2011 two definitions of gamification got published. Deterding et al. \cite{Deterding2011d} defined gamification  as \textit{``the use of game design elements in non-game contexts''}. Huotari and Hamari \cite{Huotari2012a} defined  it as \textit{``a process of enhancing a service with affordances for gameful experiences in order to support user's overall value creation''}. 
We interpret both definitions as implying a goal as the utility of gamification. Both describe elements of the game design world which could change a user's experience in a different context (non-game \cite{Deterding2011d}, service \cite{Huotari2012a}). Interestingly, for Deterding ``[...]the term `gameful design' -- design for gameful experiences -- was also introduced as a potential alternative to `gamification'''\cite{Deterding2011d}. Summarizing, in Deterding's definition the goal is rather geared towards the (improved) user experience itself, in Huotari and Hamari's definition it is the outcome driven by the user experience. 
On the basis of these definitions \textbf{we assume that an implementation of gamification is containing an, at least underlying, goal}.


\subsection{Game Design Elements}
\label{sec:gamedesignelement}
An important aspect of successful gamification should be the selection of game design elements. Game design elements determine what type of gameful experiences will be generated for the users. In \cite{Deterding2011d}, Deterding et al. provided five levels of game design elements. He distinguished between: Game interface design patterns (e.g. badges, leaderboard and level); game design patterns and mechanics (e.g. time constraint, limited resources, turns); game design principles and heuristics (e.g. enduring play, clear goals, variety of game styles); game models (e.g. MDA, challenge, fantasy, curiosity, etc.) and game design methos (e.g. playtesting, playcentric design, value conscious game design). 


\subsection{User Individuality}
\label{sec:gamedesignuser}
Designing gamification is also always a user oriented process. This is due to the fact that users are all individuals driven by different input factors like age, gender, education, social skills and cross-cultural influences \cite{Hamari2013a,Khaled2011,Yang2011,Yee2006,Yee2012}. In the game world this is considered by several player typologies developed on user observations and in-game behavior.
Hamari et al. stated existing game player typologies \cite{Hamari2014a}. They found out that player types have their legitimation because of the different behavior and motivation of players. It is a wide-spread assumption that also for the gamification scenario such types of players respectively users can be applied. Although many player typologies exist we argue that it is hard to map them to one or more specific game design elements. Beyond that, such types could change over time which seems to be a central criticism on player typologies \cite{Hamari2014a}. Furthermore, we argue that applying a set of game design elements to cover all different types in a gamification scenario could have negative influence to each other.

\subsection{Relation: User -- Game Design Element}
\label{sec:gamedesignuerelation}

``Play-personas are suggested as a useful tool that can be used to put player type research into practice as part of the design process of gamified systems.''\cite{Dixon2011} Therefore, we need to reduce the effort on determine relevant player types for implementing gamification. Why not skipping the determination of player types and directly suggest game design elements? Trying to achieve this with questionnaires and interviews would of course increase the design effort. But what if we could use a formula or tool which helps to select such game design elements based on experiences learned from user interaction data over time? We argue that this would not only reduce the design effort, furthermore this could also provide a better selection of game design elements. Because, these kind of selection would not be based only on how user perceive gamification\cite{Meder2013} but on their actual interaction with game design elements. At least such data-centric tool could support the design process substantially.
\\
User interaction with game design elements could be measured by logging activity data like mouse clicks and mouse over movements on such elements. Also the number of interface views containing specific game design elements are easy to log.
\\
For example, with a web client interface, very frequent visits of a user on a web page or web view displaying a leaderboard over points and achieved badges would indicate that the game design element leaderboard has an effect to the user. Another user might hover frequently on user interface element which displays their detailed expertise level only on mouse over. We argue that analyzing such data could help to improve the gamification design process.



\subsection{The Gamification Design Problem}
\label{sec:gamedesignselection}
Finding an optimal \textit{user} and \textit{game design elements} relation implies a goal or outcome we want to achieve with that relation. Thus, regarding the predefined goal of a gamification implementation extends the user and game design element relation to a \textit{goal, user and game design elements relation}. In this relation the right selection of game design elements is crucial to reach the goal. 
\\
\textbf{We consider the gamification design problem as the problem of assigning each user a game design element that maximizes their expected contribution to achieve some goal.}


\section{A Statistical Approach}
\label{sec:problem}
This section formalizes the gamification problem and suggests a solution based on interaction data in terms of statistical learning theory \cite{Vapnik2000}. Parts of the treatment are based on \cite{Vapnik2000,Said2012}.

\subsection{A General Model of Gamification}
We suggest a general model of gamification consisting of the following four components:
\begin{itemize}
\item A task $T$ that need to be performed.  
\item A set of game design elements $g \in \S{G}$. 
\item A set of users $u \in \S{U}$ processing task $T$ enhanced by $\S{G}$.
\item A task-dependent ground truth
\[
f_*:  \S{U} \rightarrow \S{G}.
\]
\item A function class $\S{F}$ consisting of functions of the form
\[
f: \S{U} \rightarrow \S{G}.
\]
\end{itemize}
The gamification problem is the problem of selecting a function $f \in \S{F}$ that best approximates the supervisor $f_*$.
\\
The ground truth $f_*$ is a function that assigns each user $u$ a game design element $g$ that maximizes the expected contribution of $u$ to achieve a pre-specified goal. For users that best perform without any of the game design elements contained in $\S{G}$ we include a distinguished symbol $\varepsilon$ denoting the absence of any design element. 
\\
Typically, the ground truth is unknown for most users and therefore need to be approximated by a function from some function class $\S{F}$ based on a small subset 
\[
\S{Z} = \cbrace{(u_1, g_1), \ldots (u_n, g_n)} \subseteq \S{U} \times \S{G}
\]
of training examples. The training set $\S{Z}$ consists of $n$ users $u_i$ with corresponding design elements $g_i = f_*(u_i)$  for which the ground truth is known.
\\
Note that we do not want to memorize the training examples but rather find (learn) a function $f \in \S{F}$ that predicts the best fitting design elements for new users not considered in $\S{Z}$. 

 \subsection{Learning Problem}
 
\begin{figure}[t]
\centering
\includegraphics[width=0.4\textwidth]{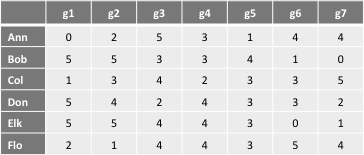}
\caption{Utility-scores for six users and seven game design elements. Scores are values from $\cbrace{0, 1, \ldots, 5}$. Higher scores indicate higher utility and vice versa.}
\label{fig:example01}
\end{figure}

\begin{figure}[t]
\centering
\includegraphics[width=0.4\textwidth]{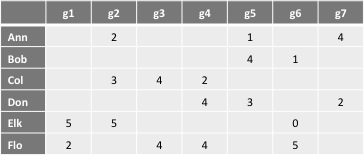}
\caption{Sparse user-design matrix of utility-scores consisting of six users and seven game design elements.}
\label{fig:example02}
\end{figure}

There are different ways to select (learn) a function $f$ from $\S{F}$ in order to approximate the ground truth $f_*$. 
\\
One approach describes users $u$ by a feature vector $\vec{x}_u$. The components of $\vec{x}_u$ measure different properties of that user such as, for example, click behavior, mouse movements and other features. Then a classifier such as the support vector machine \cite{Cortes1995} is trained to learn a model that predicts the best fitting game design element for new users.
\\
Here we consider a second approach based on user interaction with different game design elements. We measure the utility of a game design element $g$ for user $u$ in achieving task $T$ by means of a utility function
\[
f_U: \S{U} \times \S{G} \rightarrow \R, \quad (u, g) \mapsto s_{ug}.
\]
The utility-scores $s_{ug}$ capture to which extent each user $u$ together with design element $g$ contributes to some overall goal. Given a utility function $f_U$, we select a classifier $f \in \S{F}$ according to the rule
\[
f(u) = g_u^* = \arg\max_{g \in \S{G}} f_U(u, g).
\]
Thus, $f$ assigns user $u$ a game design element $g_u^*$ with maximum utility.
\\
Figure \ref{fig:example01} provides an example of a utility function $f_U$ shown in matrix form $\vec{S} = (s_{ug})$. In this example, we would assign game design $g_3$ to user \emph{Ann}. For user \emph{Bob} the maximum score of $5$ is achieved for design elements $g_1$ and $g_2$. In this case, we can pick either $g_1$ or $g_2$ as design element for \emph{Bob}.
\\
In practice, however, the matrix $\vec{S}$ is sparse for various reasons. For example, users might no be willing to explore all design elements and may quit using the system. Figure \ref{fig:example02} provides an example for the case of a sparse matrix $\vec{S}$ of utility-scores. In this scenario, we aim at learning $f_U$ on the basis of $n$ observations $(u_1, g_1, s_1), \ldots, (u_n, g_n, s_n) \in \S{U} \times \S{G} \times \R$ consisting of $n$ users $u_i$ together with corresponding game design elements $g_i$ and utilitiy-scores $s_{i}$. 
 
The problem of gamification reduces to estimating a functional relationship 
\[
f: \S{U} \times \S{G} \rightarrow \R, \quad (u,g) \mapsto \hat{s}_{ug}
\]
that \emph{best} predicts the utility-score $s_{ug}$ of design element $g$ for user $u$ by means of $f(u, g) = \hat{s}_{ug}$. To clarify what we mean by \emph{best}, we introduce the notion of loss function. A loss function $\ell(\hat{s}, s)$ measures the cost for predicting $\hat{s}$ when the true utility-score is $s$.  A common choice for a loss function is the squared error loss defined by
\[
\ell(\hat{s}, s) = \args{\hat{s}-s}{^2}.
\]
Our goal is to find a function that minimizes the expected loss 
\[
E[f] = \int \ell(f(u, g), s_{ug}) dP(u, g, s_{ug})
\]
where $P(u, g, s)$ denotes the joint probability distribution on $\S{U} \times \S{G} \times \R$.

Suppose that we know a function (ground truth) $f_*$  that minimizes the expected loss $E[f]$. Then we are in a similar situation as in the above scenario, where each user has explored all game design elements. The complete user-design matrix $\vec{S} = (s_{ug})$ has elements of the form
\[
s_{ug} = f_*(u,g).
\] 
We can assign each user $u$ a game design element $g_{u}^*$ according to the following rule
\[
g_{u}^* = \arg max_{g} f_*(u,g).
\]
In practice, we neither know $f_*$ nor the joint probability distribution $P(u, g, s_{ug})$. Therefore we can not find a minimum $f_*$ of $E[f]$ directly. Instead  we try to approximate $f_*$ by a function $\hat{f}_*$ that minimizes the empirical loss 
\begin{align*}\label{eq_p1}
\hat{E}[f] &= \frac{1}{n}\sum_{i=1}^n \ell(f(u, g), s_{ug}).
\end{align*}
on the basis of a sample of observed data 
\[
(u_1, g_1, s_1), \ldots, (u_n, g_n, s_n).
\]
The sparse user-design matrix shown in Figure \ref{fig:example02} is an example of a sample of observed data.
\\
According to the empirical risk minimization principle, this approach is statistically consistent \cite{Vapnik2000}, meaning that the approximation $\hat{f}_*$ converges to the true minimum $f_*$ with increasing amount of data. The learning problem consists in predicting the missing values.
\\
This setting reduces the gamification problem of finding the best design element for each user to the problem of regression learning for which a plethora of powerful mathematical methods are available.

\subsection{Matrix Completion}
\label{sec:relationship}
\begin{figure}[t]
\centering
\includegraphics[width=0.4\textwidth]{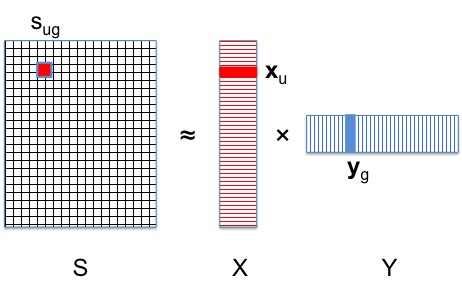}
\caption{Matrix factorization. Approximate the user-design matrix $S$ by low-rank matrices $X$ and $Y$.}
\label{fig:mf}
\end{figure}

The gamification problem as proposed in this section can be regarded as a special case of a recommendation problem \cite{Kantor2011} for which matrix factorization constitutes a state-of-the-art solution \cite{Candes2009,Koren2009,Lee1999}.  

Matrix factorization characterizes users and game design elements by $k$ factors (properties) inferred from the utility-score patterns hidden in the user-design matrix $S = (s_{ug})$. Users $u$ and game design elements $g$ are associated with vectors $\vec{x}_u \in \R^k$ and $\vec{y}_g \in \R^k$, respectively. The $k$ elements in $\vec{y}_g$ measure to which extent design element $g$ possesses these factors.  Similarly, the elements in $\vec{x}_u$ measure to which extent user $u$ prefers these factors.  High correspondence between factors of user $u$ and factors of design element $g$ indicate high utility. Correspondence between user and design factors is modelled as inner product such that 
\begin{align}\label{eq:p=xTy}
s_{ug} \approx \vec{x}_u^T\vec{y}_g
\end{align}
for all known utility-scores $s_{ug}$. In matrix notation, eq.\ \eqref{eq:p=xTy} takes the form
\[
S \approx X\cdot Y,
\]
where $X$ is the user matrix and $Y$ is the game design element matrix. The rows  $\vec{x}_u^T$ of $X$ and the columns $\vec{y}_g$ of $Y$ describe the users $u$ and design elements $g$, respectively. Figure \ref{fig:mf} illustrates how the user-design matrix $S$ is factorized by low-rank matrices $X$ and $Y$.

\begin{figure}[t]
\centering
\includegraphics[width=0.4\textwidth]{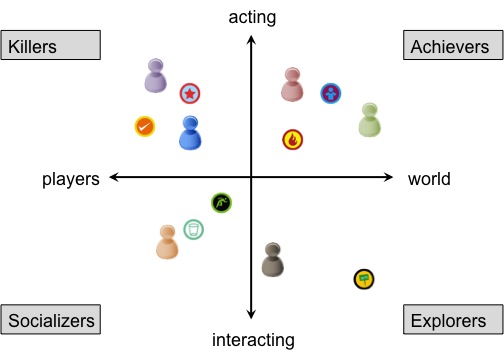}
\caption{A simplified illustration of the latent factor space generated by matrix factorization. The latent factors refer to the preferences indicated by the $x$- and $y$-axis. The six users and seven design elements of Figure \ref{fig:example01} are embedded into the factor space. According to Bartle's player typology \cite{Bartle1996}, users fall into one of the four categories \emph{achiever, explorer, socializer}, and \emph{killer}. Similarly, the features of the design elements refer to characteristics of the user categories.}
\label{fig:bartle}
\end{figure}

Figure \ref{fig:bartle} shows an fictitious example of how the six users and seven game design elements from Figure \ref{fig:example02} are associated to vectors from the two-dimensional vector space $\R^2$. The latent factors are inferred from the utility-score patterns hidden in the user-design matrix $S$. In this example, the two discovered factors refer to the preferences according to the player typology proposed by Bartle \cite{Bartle1996}. In practice, however, there may be additional $(k > 2)$ or different factors, which may be completely uninterpretable for humans. 

After all users and all game design elements have been embedded into the joint latent factor space $\R^k$, missing values $s_{ug}$ of the sparse matrix $S$ can be predicted in a straight forward way by 
\[
\hat{s}_{ug} = \vec{x}_u^T\vec{y}_g
\]
in order to complete matrix $S$. 

To learn the embeddings into the factor space $\R^k$, we need to solve the following basic problem
\begin{align*}
(P) &:\qquad \min_{\vec{x}_*, \vec{y}_*}   \sum_{(u,g) \in \S{P}} \argsS{s_{ug}-\vec{x}_u^T\vec{y}_g}{^2},
\end{align*}
where $\S{P}$ is the set of all pairs $(u,g)$ for which $s_{ug}$ is known.

\section{Discussion}
Under the assumption that different users experience the same game design elements differently, we define the gamification problem as the problem of assigning each user a game design element such that their expected contribution to achieve some pre-specified goal is maximized. 
\\
One way to assign design elements to users is by means of customer segmentation. In marketing theory, segmentation aims at identifying customer groups in order to better match the needs and wants of customers. For games those customer segments correspond to different player types based on character theory. Once a user is classified into a customer segment, an appropriate design element for that segment is selected and assigned to that user. The hardest part of this approach is to design categories that correspond to various dimensions describing characteristic features of users such as the multiple motivations of varying degrees existing simultaneously across users and user types. 
\\
To avoid assignments of design elements to users via the indirection of customer segments and user types from marketing and character theory, resp., we aim at learning a predictive model based on statistical principles that directly classifies users to game design elements. Based on user interaction with game design elements, we suggest to solve the learning problem by means of matrix factorization. The latent factors discovered by a matrix factorization model may be interpreted as characteristic properties of game design elements. User factors describe to which extent a user prefers such characteristic features. Thus, the latent factors can be regarded as a computerized alternative to the aforementioned customer segments and user types.
\\
To keep the gamification model simple, we ignored time dynamics of user preferences leaving this issue open for future research. In addition, learning classifiers based on user behavior characteristics is a second issue for future research. The main challenges consist in constructing a useful utility function when using matrix factorization and generating useful behavior features when learning classifiers. Due to lack of publicly available data,  empirical evaluations are currently not possible. Therefore this contribution aims at directing the design process of gamification to a more principled way based on statistical grounds.  

\commentout{
\section{Acknowledgments}
This section is optional; it is a location for you
to acknowledge grants, funding, editing assistance and
what have you.  In the present case, for example, the
authors would like to thank Gerald Murray of ACM for
his help in codifying this \textit{Author's Guide}
and the \textbf{.cls} and \textbf{.tex} files that it describes.
}

%
\bibliographystyle{abbrv}
\bibliography{sigproc}  

\begin{thebibliography}{10}

\bibitem{Anderson2013}
A.~Anderson, D.~Huttenlocher, J.~Kleinberg, and J.~Leskovec.
\newblock {Steering user behavior with badges}.
\newblock In {\em Proc. WWW}, pages 95--106, Rio de Janeiro, Brazil, 2013.

\bibitem{Bartle1996}
R.~Bartle.
\newblock {Hearts, clubs, diamonds, spades: Players who suit MUDs}.
\newblock {\em Journal of MUD research}, 1(1):19, 1996.

\bibitem{Candes2009}
E.~J. Cand\`{e}s and B.~Recht.
\newblock {Exact Matrix Completion via Convex Optimization}.
\newblock {\em Foundations of Computational Mathematics}, 9(6):717--772, 2009.

\bibitem{Cheng2011a}
L.-T. Cheng, S.~Shami, C.~Dugan, M.~Muller, J.~DiMicco, J.~Patterson, S.~Rohal,
  A.~Sempere, and W.~Geyer.
\newblock {Finding Moments of Play at Work}.
\newblock In {\em Workshop on Gamification: Using Game Design Elements in
  Non-Gaming Contexts}, pages 2--5, 2011.

\bibitem{Cortes1995}
C.~Cortes and V.~Vapnik.
\newblock Support-vector networks.
\newblock {\em Machine learning}, 20(3):273--297, 1995.

\bibitem{Deterding2011d}
S.~Deterding, D.~Dixon, R.~Khaled, and L.~Nacke.
\newblock From game design elements to gamefulness: defining gamification.
\newblock {\em Proceeding of the 15th International Academic MindTrek
  Conference}, pages 9--15, 2011.

\bibitem{Dixon2011}
D.~Dixon.
\newblock {Player types and gamification}.
\newblock In {\em In Workshop on Gamification at CHI2011}, pages 12--15, 2011.

\bibitem{Farzan2008a}
R.~Farzan and J.~DiMicco.
\newblock {When the experiment is over: Deploying an incentive system to all
  the users}.
\newblock In {\em Persuasive Technology}. ACM, 2008.

\bibitem{Farzan2008}
R.~Farzan, J.~DiMicco, and D.~Millen.
\newblock {Results from deploying a participation incentive mechanism within
  the enterprise}.
\newblock In {\em Proceedings of the SIGCHI conference on Human factors in
  computing systems}, pages 563--572, 2008.

\bibitem{Hamari2013}
J.~Hamari.
\newblock {Transforming Homo Economicus into Homo Ludens: A Field Experiment on
  Gamification in A Utilitarian Peer-to-Peer Trading Service}.
\newblock {\em Electronic Commerce Research and Applications}, (12), 2013.

\bibitem{Hamari2013a}
J.~Hamari and J.~Koivisto.
\newblock {Social motivations to use gamification: An empirical study of
  gamifying exercise}.
\newblock {\em Proc. ECIS'13}, pages 1--12, 2013.

\bibitem{Hamari2014}
J.~Hamari, J.~Koivisto, and H.~Sarsa.
\newblock Does gamification work? - a literature review of empirical studies on
  gamification.
\newblock In {\em In proceedings of the 47th Hawaii International Conference on
  System Sciences}, 2014.

\bibitem{Hamari2014a}
J.~Hamari and J.~Tuunanen.
\newblock {Player Types: A Metasynthesis}.
\newblock In {\em Transactions of the Digital Games Research Association},
  2014.

\bibitem{Huotari2012a}
K.~Huotari and J.~Hamari.
\newblock Defining gamification: a service marketing perspective.
\newblock {\em Proceeding of the 16th International Academic MindTrek
  Conference}, pages 17--22, 2012.

\bibitem{Kantor2011}
P.~B. Kantor, L.~Rokach, F.~Ricci, and B.~Shapira.
\newblock {\em {Recommender Systems Handbook}}.
\newblock Springer, 2011.

\bibitem{Khaled2011}
R.~Khaled.
\newblock {It's Not Just Whether You Win or Lose: Thoughts on Gamification and
  Culture}.
\newblock In {\em Workshop on Gamification: Using Game Design Elements in
  Non-Gaming Contexts}, pages 1--4, 2011.

\bibitem{Koren2009}
Y.~Koren, R.~Bell, and C.~Volinsky.
\newblock {Matrix Factorization Techniques for Recommender Systems}.
\newblock {\em IEEE Computer}, 42(8):42--49, 2009.

\bibitem{Lee1999}
D.~D. Lee and H.~S. Seung.
\newblock {Learning the parts of objects by non-negative matrix factorization}.
\newblock {\em Nature}, 401(6755):788--91, 1999.

\bibitem{Meder2013}
M.~Meder, T.~Plumbaum, and F.~Hopfgartner.
\newblock Perceived and actual role of gamification principles.
\newblock In {\em First Workshop on Crowdsourcing and Gamification in the Cloud
  (CGCloud) in conjunction with the 6th IEEE/ACM International Conference on
  Utility and Cloud Computing}, 2013.

\bibitem{Mosca2012}
I.~Mosca.
\newblock {+10! Gamification and deGamification}.
\newblock {\em G|A|M|E}, 1(1), 2012.

\bibitem{Nikkila2011}
S.~Nikkila, S.~Lin, H.~Sundaram, and A.~Kelliher.
\newblock {Playing in Taskville: Designing a Social Game for the Workplace}.
\newblock In {\em Workshop on Gamification: Using Game Design Elements in
  Non-Gaming Contexts}, pages 1--4, 2011.

\bibitem{Said2012}
A.~Said, B.~Jain, S.~Narr, and T.~Plumbaum.
\newblock Users and noise: The magic barrier of recommender systems.
\newblock In {\em User Modeling, Adaptation, and Personalization}, pages
  237--248. Springer, 2012.

\bibitem{Vapnik2000}
V.~Vapnik.
\newblock {\em {The Nature of Statistical Learning Theory}}.
\newblock Springer, 2000.

\bibitem{Yang2011}
J.~Yang, M.~R. Morris, J.~Teevan, L.~A. Adamic, M.~S. Ackerman, and O.~M. Way.
\newblock {Culture Matters : A Survey Study of Social Q \& A Behavior}.
\newblock In {\em International AAAI Conference on Weblogs and Social Media},
  pages 409--416, 2011.

\bibitem{Yee2006}
N.~Yee.
\newblock {Motivations for play in online games}.
\newblock {\em CyberPsychology \& Behavior}, pages 772--775, 2006.

\bibitem{Yee2012}
N.~Yee, N.~Ducheneaut, and L.~Nelson.
\newblock {Online gaming motivations scale: development and validation}.
\newblock {\em Proc. CHI '12}, pages 2803--2806, 2012.

\end{thebibliography}
%
%

\end{document}